\documentstyle[aps,prl,epsfig]{revtex}

\begin{document}
\twocolumn[\hsize\textwidth\columnwidth\hsize\csname@twocolumnfalse\endcsname
\title{Intrinsic decoherence and classical-quantum correspondence
in two coupled delta-kicked rotors}
\author{Hwa-Kyun Park\footnote{e-mail: childend@complex.kaist.ac.kr \\
 TEL:+49-351-871-1225 \hspace{.5cm} FAX: +49-351-871-1999}, Sang Wook Kim}
\address{Max-Planck-Institut f{\"u}r Physik komplexer Systeme,
N{\"o}thnitzer Str. 38, D-01187 Dresden, Germany}
\date{\today}

\maketitle

\begin{abstract}
We show that classical-quantum correspondence of center of mass motion in two 
coupled delta-kicked rotors can be obtained from intrinsic decoherence of the 
system itself which occurs due to the entanglement of the center of mass motion 
to the internal degree of freedom without coupling to external environment.
\end{abstract}

\pacs{PACS number(s): 05.45.Mt, 03.65.Yz, 03.65.Sq}
\narrowtext
\vskip1pc]

Classical-quantum correspondence in a classically chaotic system has been one of 
the most interesting problem in physics for a long time \cite{reichl,zurek91}. 
In quantum mechanics, the time evolution of a wave function follows a {\em linear} 
Schr\"odinger equation, and so there is no possibility of a sensitivity on the 
initial condition, a trademark of classical chaos. Also, chaotic diffusion is 
suppressed by quantum localization \cite{casati79}. It has been revealed that some 
crossover time $t_{r}=\log(I/\hbar)/\lambda$ ($I$ is a characteristic action and 
$\lambda$ is a Lyapunov exponent) exists so that classical-quantum correspondence 
breaks down for $t>t_r$ \cite{time}. As $\hbar \rightarrow 0$ the crossover time 
increases indefinitely and classicality is recovered. However, the problem is that 
$\hbar$ is a nonzero constant and $t_r$ is not sufficiently large considering its 
logarithmic dependence of $\hbar$ \cite{zurek94}.

Recently, the relation between decoherence and the classical-quantum correspondence 
has been investigated extensively 
\cite{zurek91,zurek94,zurek93,habib98,Pattanayak99,miller99,paz00,nag01,miller99b}.
Decoherence breaks the purity of initial superposition, which should 
be conserved in the absence of coupling to the environment, and thus only the partial 
fraction of whole Hilbert space, namely pointer states, are selected by the environment 
\cite{zurek93}. The dynamics of the system coupled to the environment shows the 
unique characteristics of the system independent of the coupling strength as long 
as it is not too large or small \cite{paz00}. In other words, with appropriate 
coupling to environment, the Lyapunov exponent or entropy production rates, which are 
important physical quantities characterizing a chaotic system, can be reproduced 
quantum mechanically. 
However, there has been some debate on whether decoherence from environment
is indispensable to obtain the classical-quantum correspondence for classically 
chaotic systems \cite{zurek94}. 

In this letter, we show that decoherence can occur naturally in composed system 
even when we ignore the coupling to the outer environment. The initially pure center 
of mass states become dynamically entangled to the internal degrees of freedom, 
which effectively acts like the environment. Let us consider a classical object 
governed by a Hamiltonian $H_{0}=P^{2}/2M + V(X)$, where $M$ is the mass of the classical
object.
Since classical objects are composed of many particles,
a complete Hamiltonian will be given by 
$H=\sum [p_{i}^2/2m_i + V_{1}(x_{i})]+ V_{2}(x_{1},x_{2},\cdots)$, where
$p_{i}$ and $x_{i}$ are momenta and coordinates of the constituents, respectively. 
We assume that the mass of each constituent $m_i$ is equal to $m$. If the force 
$f_{i}=-d V_{1}(x_{i})/dx_{i}$ is linear, we can ignore the motion of the individual 
constituent particles to describe the center of mass dynamics of the macroscopic classical
object since $M\ddot{X}$=$\sum f_{i}= k \sum x_{i} = NkX$, where $X$, $N$, and $k$ are 
a position of center of mass, the number of constituents, and a constant characterizing the 
linear force, respectively. If the force $f_i$ is nonlinear, however, the description of 
the motion of the classical object by using only $H_0$ is not immediately clear. 
We already know from our experience, nevertheless, that the classical dynamics 
of the macroscopic object is well described by the Hamiltonian $H_{0}$. 
It means that $\sum f_{i}$ should almost coincide with $F(X)=-dV(X)/dX$ in 
values. Thus, we can ignore the difference between $H$ and $H_0$ to consider the 
motion of classical objects. However, the minor differences between $H$ and $H_{0}$ 
play an important role in quantum mechanics since the internal degree of freedom can 
induce decoherence on the center of mass motion. In this case, the correct 
classical-quantum correspondence of the center of mass motion will be obtained not 
from $H_0$ but from $H$.

Only due to computational difficulty, in this letter we consider a two particle 
system. 
Even with this simple model, we have observed some evidence of the {\em intrinsic}
decoherence originating from the entanglement to internal dynamics. We will show
the delocalization of wavefunctions in momentum space and the coincidence of classical 
and quantum entropy production rates. 
In the previous study of two coupled quantum kicked tops, the entanglement of two 
initially decoupled kicked tops increases linearly in time with a rate which is 
a linear function of the sum of the positive Lyapunov exponents \cite{miller99b}.   
However, the classical limit has not been taken, and thus the classical-quantum 
correspondence has not been considered although signature of classical chaos has 
been observed for low quantum numbers. A model of two 
interacting spins was also studied to investigate the correspondence of 
classical and quantum Liouville dynamics\cite{emerson01}. But, 
this correspondence was considered to have nothing to do with 
the decoherence effect.

The main purpose of this letter is to elucidate that the intrinsic decoherence due to the 
entanglement of subsystems plays an important role to understand the classical-quantum 
correspondence of composed systems. For this purpose, we consider the center of mass 
motion of the delta-kicked rotors composed of two particles of which a governing 
Hamiltonian is given by
\begin{eqnarray}\label{h0}
\lefteqn{H=\frac{1}{2m} (p_{1}^2 + p_{2}^2) + U(r_{1},r_{2})} \hspace{2cm}  \nonumber \\
  &   & + k [\cos (r_{1})+\cos(r_{2})] \sum_{i} \delta (t-i T),
\end{eqnarray}
where $r_1$ and $r_2$ are angle variables with range $2\pi$ radians and $U(r_{1},r_{2})$ 
is the interaction potential which confines two particles within a distance $w$.
If the confinement width $w$ goes to zero, then the system reduces to a usual delta-kicked 
rotor.

To investigate the center of mass motion, we introduce canonical transforms,
$R=(r_{1}+r_{2})/2$, $r=r_{1}-r_{2}$, $P=p_{1}+p_{2}$, and $p=(p_{1}-p_{2})/2$.
Then, we obtain the following Hamiltonian,
\begin{equation}\label{hcn}
H=\frac{P^{2}}{2M} + \frac{p^{2}}{2\mu} + U(r) 
+ K \cos (R) \cos \left( \frac{r}{2} \right) \sum_i \delta (t-i T),
\end{equation}
where $M=2m$, $\mu=m/2$, and $K=2k$. Here, $U(r)$ corresponds to the confining potential
with impenetrable walls at $r=\pm w$. Delta-kicks described by the last term of 
Eq.~(\ref{hcn}) yield the interaction between the center of mass motion and the 
internal degree of freedom, i.e. the motion of the reduced mass $\mu$. 

The time evolution of a wave function $\Psi(R,r,t)$ is given by simple maps. Between each kick 
occurring at $t=iT$ ($i$ is an integer), the center of mass motion and the internal motion
evolve independently, so that we obtain
\begin{eqnarray}\label{map1}
\lefteqn{\psi(R,r,NT+0^{-})=\exp\left(-i \frac{P^{2}}{2M} \frac{T}{\hbar} \right)}\hspace{0.5cm}\nonumber \\
                   & & \exp\left\{-i\left[\frac{p^{2}}{2\mu} + U(r)\right] \frac{T}{\hbar}\right\}
			 \Psi(R,r,(N-1)T+0^{+}).
\end{eqnarray}
The wave functions just before and after the delta-kick at $t=NT$ are related by 
the following mapping:
\begin{eqnarray}\label{map2}
\lefteqn{\psi(R,r,NT+0^{+}) = } \hspace{0.5cm}  \nonumber \\
  & & \exp \left[ -i K \cos (R) \cos \left( \frac{r}{2} \right) \right] \Psi(R,r,NT+0^{-}).
\end{eqnarray}
Combining two maps given in Eq.~(\ref{map1}) and (\ref{map2}), we numerically calculate the 
evolution of the wave function $\Psi(R,r,t)$ for various $\hbar$ and $w$. For most cases, the 
number of basis states used for the motion of coordinate $R$ and $r$ are 16384 and 256 respectively, 
while 32768 and 512 basis are used respectively for small $\hbar$. The 
initial condition is chosen to be the ground state $\Psi(R,r,0)=1/\sqrt{w} \cos(\pi r/2w)$. 

Classical evolution is obtained from a four dimensional map for $R,P,r$ and $p$ derived 
from Hamiltonian (\ref{hcn}). We consider an ensemble of particles which are  distributed 
from $R=0$ to $R=2\pi$ uniformly with $P=0$, and from $r=-w$ to $r=w$ with a probability 
$\cos^2(\pi r/2w)/w $ with $p=\pm p_{0}=\pm (\pi \hbar)/(2w)$. From this initial condition, 
the classical evolution is simulated, and the ensemble average of the normalized 
variance of center of mass momentum, 
$\Delta^2=2\left(\left<P^2\right>-\left<P\right>^2\right)/MK^2$ is computed. 

As a reference, let us mention the case with a single delta-kicked rotor governed by the 
Hamiltonian $H = P^2 /2M  +  K \cos (R) \sum_{i} \delta (t-i T)$, which corresponds to the case with
$w=0$.  For $K=5$, the system is fully chaotic, and classical 
kinetic energy increases diffusively. But the diffusion is suppressed quantum mechanically, 
and the classical-quantum correspondence breaks down, which is the well-known
dynamical localization \cite{casati79}.

Now, we consider two delta-kicked rotors.
The inset in Fig.~1 shows the differences of classical momentum 
variances between single and two delta-kicked rotors for various $w$.
As $w$ is decreased, the difference between them vanishes. 
Meanwhile, shown in Fig.~1 are the differences 
between the classical and the quantum variances for various $w$ with $\hbar=0.07$.
One clearly sees that the quantum localization breaks down when w is increased.
The difference of $\Delta_{cl}$ and $\Delta_{qm}$ nearly 
disappears for large $w$, while deviations are observed for small $w$. 
The break-down of quantum localization, i.e., delocalization, is directly visible in the 
momentum distribution function in Fig.~2, where the probability distribution of the center 
of mass momentum $P$ are shown at $n=500$ for several $\hbar$ and $w$. The exponential
localization is changed into a rather broad Gaussian-like profile as we decrease $\hbar$ 
or increase $w$. Let us note that the delocalization of wavefunctions alone is not enough to
prove the occurrence of decoherence. In fact, the delocalization was also observed in the
study of two interacting particles (TIP) in a random potential\cite{tip1,tip2}. 
However, it was not attributed to the decoherence. 

Next, we consider the reduced density matrix for the center of mass motion in order to show 
that the observed classical-quantum correspondence is ascribed to the intrinsic decoherence 
caused from the interaction with the internal degree of freedom. The reduced density matrix 
$\rho_R(R_{1},R_{2})$ is given by $Tr_r(\rho)=\sum_{r} \psi(R_{1},r) \psi^{*}(R_{2},r)$. 
As a quantitative measure of decoherence, we calculate the linear entropy 
$s_{l}=Tr(\rho_{R} - \rho_{R}^2)$ \cite{zurek93}. Note that for a pure state $s_{l}=0$, while 
for a maximum decoherence $s_{l}=1$. Figure 3 shows that for large $w$ the entropy $s_{l}$ 
rapidly approaches 1, i.e., a maximum decoherence. As we decrease $w$, the entropy $s_{l}$ 
shows rather reduced values and slowly increases in time. In fact, an energy level spacing 
of the internal dynamics described by $p^2/2\mu + U(r)$ is proportional to $1/w^2$, which
means the smaller $w$, the larger the level spacing. For a given perturbation determined by $K$
it is more difficult to excite the internal dynamics in the case of large spacing,
which leads to effective decoupling between the center of mass motion and the internal
degree of freedom, and eventually to the decrease of decoherence. This is consistent with 
the previous result that the break-down of the localization and the classical-quantum 
correspondence is easily obtained for large $w$, i.e., strong decoherence. 

The $\cos(r/2)$ in the last term of Hamiltonian (\ref{hcn}) can be regarded as an amplitude 
noise of the kick onto the center of mass motion. As $w$ decreases, the noise is reduced 
and so is the effect of decoherence. Ott {\it at al.} studied the effect of noise on the
single delta-kicked rotor, which showed that, for moderate noise and $\hbar$,
the diffusion coefficient $D_{qm}$ is proportional to the square of the noise level, 
more precisely $D_{qm} \sim \nu ^{2} (K/\hbar)^{4}$ ($\nu$ is a noise strength) \cite{ott84}. 
For very small $\hbar$, it is obtained that $D_{qm} \simeq D_{cl}$ ($D_{cl}$ is a classical diffusion 
coefficient). If we directly apply this result to our case, we immediately obtain the scaling 
relation $D_{qm} \sim (w K/\hbar)^{4}$ for moderate $\hbar$ since we can approximately 
regard $1-\cos(r/2)$ as a noise amplitude, and so $\nu \sim w^2$.
This scaling is confirmed by numerically calculating $D_{qm}$ as a function of $(w K/\hbar)$ 
in Fig.~4, 
where $\hbar=0.25$, and $K=5$. These results also confirm the above proposition that 
the term $\cos(r/2)$ in the Hamiltonian (\ref{hcn}) can be treated as noise and thus induces
decoherence.

Finally, we consider the classical and the quantum entropy production rates of which coincidence 
has been the criteria for the classical-quantum correspondence of chaotic systems in studies of 
decoherence \cite{paz00}.   
Quantum mechanically, the von Neumann entropy is given by $S_{vn}=Tr[\rho_R \log(\rho_{R})]$. Due to the 
difficulty of computing the eigenvalues of large matrix, we use a smaller number of basis states than
that used in the previous simulation, 4096 for the motion of $R$, and 256 for $r$. 
We choose $\hbar=0.07$ and can obtain the quantum evolution up to $n=100$ with this smaller number 
of basis. The time evolution of the von Neumann entropy consists of two different regimes \cite{comm1} 
as shown in Fig.~5. After the initial transient, the von Neumann entropy increases logarithmically on 
time, which exactly corresponds to its classical counterpart \cite{nag01}. Let us consider the phase 
space $(R,P)$ for the center of mass motion. Classical distributions are uniform in $R$ and Gaussian 
in $P$. The variance of momentum increases diffusively.  As a result, 
the classical entropy is given by $S_{cl} \sim 0.5 \log(n)$. 
In Fig.~5, the quantum 
results (dots) show $0.5 \log(n)$ dependence, consistent with the classical prediction.

In summary, we have studied the dynamics of the center of mass motion of two coupled delta-kicked 
rotors and examined its classical-quantum correspondence using the momentum distribution and the 
entropy production rate. We show that the correspondence is obtained through intrinsic decoherence
which comes from the interaction between the center of mass motion and the internal degree of freedom. 
Neither external noise nor an external environment are assumed, here the decoherence results from the
internal dynamics of the system itself. 
This decoherence effect should also be observed in systems composed of many particles.
Although our study is motivated by the dynamics of the 
classical object composed of many constituents, it might also be relevant for molecules composed 
of two or many atoms. We hope that similar results would be observed in experiments using 
molecules when they are subjected to an external nonlinear force depending on the position of the constituent 
atoms.


\begin{figure}
{\epsfig{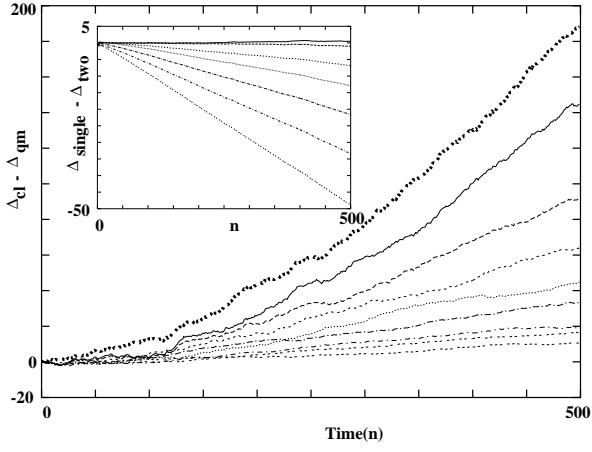}\\}
\caption{ 
The differences between classical and quantum momentum variances, 
$\Delta_{cl}-\Delta_{qm}$, are plotted with
$M=1$, $\mu=0.25$, $K=5$, $T=1$, and $\hbar=0.07$. 
From top to bottom, $w=0.0$ (i.e. single kicked rotor), $0.1,0.2,0.3,0.4,0.5,0.6$
and $0.7$. 
(Inset) The difference between variances of single and two coupled delta-kicked 
rotors, $\Delta_{single}-\Delta_{two}$ for various $w$ are shown.
From top to bottom, $w=0.1,0.2,0.3,0.4,0.5,0.6,0.7$ and $0.8$.
}
\label{p2}
\end{figure}

\begin{figure}
{\epsfig{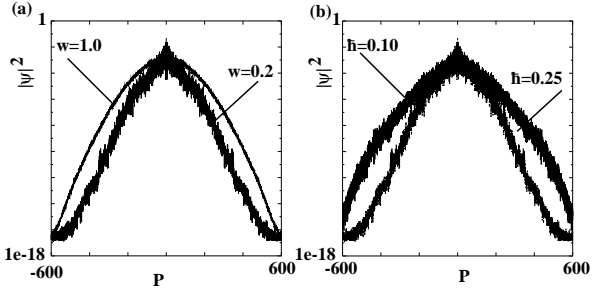}\\}
\caption{ Momentum distribution functions. 
(a) $\hbar=0.25$, $w=0.2$ and $1$. (b) $w=0.2$, $\hbar=0.1$
and $\hbar=0.25$.  Other parameters are the same as in Fig. 1. }
\label{pdf}
\end{figure}

\begin{figure}
{\epsfig{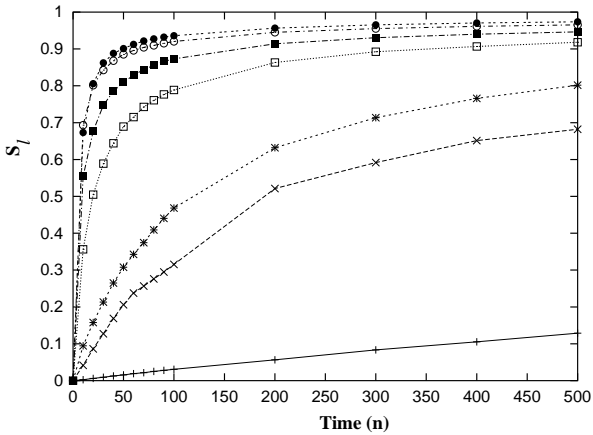}\\}
\caption{ Linear entropies for various $w$.
From bottom to top, $w=0.1,0.2,0.3,0.4,0.5,0.6$
and $0.7$. 
Other parameters are the same as in Fig. 1.
}
\label{sl}
\end{figure}

\begin{figure}
{\epsfig{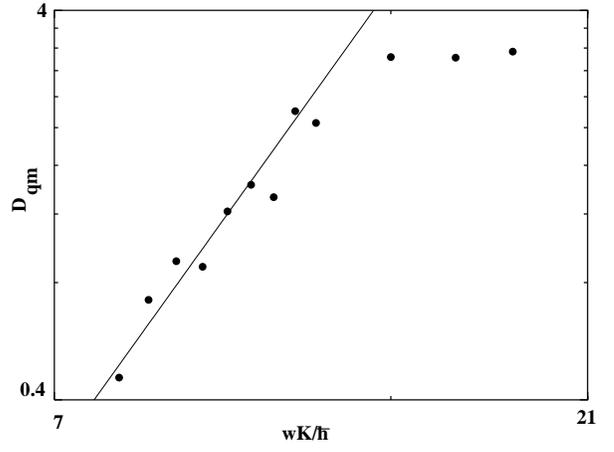}\\}
\caption{ Quantum diffusion coefficient $D_{qm}$ as a function 
of $(w K/\hbar)$ with $\hbar=0.25$. 
The solid line corresponds to $D_{qm} \propto (w k/\hbar)^4$.
Other parameters are the same as in Fig. 1.
}
\label{qd}
\end{figure}

\begin{figure}
{\epsfig{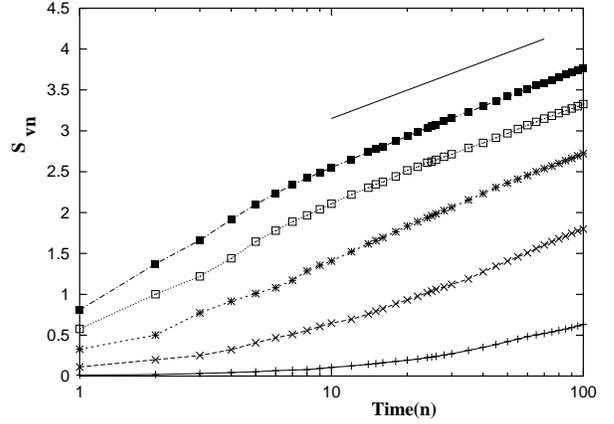}\\}
\caption{  
The von Neumann entropies obtained from the reduced 
density matrix $\rho_{R}$ with $\hbar=0.07$. For a reference,
the $0.5 \log(n)$ dependence of $S_{cl}$ is represented by the solid line.
From top to bottom, $w=0.2,0.4,0.6,0.8$ and $1.0$.   
}
\label{vns}
\end{figure}


\begin{thebibliography}{99}
\bibitem{comm1} 
Similar results have been observed in the study of the driven double well
system coupled to the external heat bath\cite{paz00}. In this case, the entropy 
production rate becomes independent of the diffusion caused by the 
environment only after some initial transient.
\end{thebibliography}
\end{document}